\documentclass[aps,prl,superscriptaddress,english,10pt,noshowpacs,nofootinbib,twocolumn]{revtex4-1}
\usepackage{graphicx}
\usepackage{float}
\usepackage{latexsym}
\usepackage{graphicx}
\usepackage{amssymb}
\usepackage{amsmath}
\usepackage{amsfonts}
\usepackage{dcolumn}
\usepackage{color}
\usepackage{xcolor}
\usepackage{amsmath}	
\usepackage{amssymb}	
\usepackage{hyperref} 	
\usepackage[all]{hypcap}
\usepackage[title]{appendix}
\usepackage{notoccite}
\usepackage[normalem]{ulem}
\hypersetup{
    colorlinks=true,	
    linkcolor=red,		
    citecolor=blue,		
    filecolor=magenta,	
    urlcolor=black	    
}
\newcommand{\dd}{\mathrm{d}}

\newcommand{\eref}[1]{Eq.~(\ref{#1})}

\newcommand{\fref}[1]{Fig.~\ref{#1}}

\newcommand{\nn}{\nonumber \\}
\def\be{\begin{equation}}
\def\ee{\end{equation}}
\def\bea{\begin{eqnarray}}
\def\eea{\end{eqnarray}}

\begin{document}
\title{Runaway gravitational production of dark photons}

\author{Christian Capanelli}
\email{christian.capanelli@mail.mcgill.ca}
\affiliation{Department of Physics \& Trottier Space Institute,
McGill University, Montr\'eal, QC H3A 2T8, Canada.}

\author{Leah Jenks}
\email{ljenks@uchicago.edu}
\affiliation{Kavli Institute for Cosmological Physics, The University of Chicago, Chicago, IL 60637}

\author{Edward W. Kolb}
\email{Rocky.Kolb@uchicago.edu}
\affiliation{Kavli Institute for Cosmological Physics and Enrico Fermi Institute, The University of Chicago, Chicago, IL 60637}

\author{Evan McDonough}
\email{e.mcdonough@uwinnipeg.ca}
\affiliation{Department of Physics, University of Winnipeg, Winnipeg MB, R3B 2E9, Canada}

\begin{abstract}
We demonstrate that gravitational particle production (GPP) of a massive, Abelian, vector (Proca) field during inflation in the presence of nonminimal coupling to gravity may suffer from an instability which leads to runaway production of high-momentum modes. This is untenable unless there is some mechanism to regulate the runaway. We discuss the parameter space of the particle mass and nonminimal couplings where such a runaway occurs and possible ways to tame the runaway. We find that there is no obvious way to resolve the runaway in a UV completion or with kinetic mixing to the standard model.
\end{abstract}

\maketitle

{\it Introduction:}  
Dark matter is known to exist and yet the identity of the dark matter particle(s) remains elusive. A natural possibility, in light of the tremendously successful theory of electroweak symmetry breaking, is that dark matter includes a massive vector field, analogous to the $W$ and $Z$ bosons of the Standard Model (SM),which interacts weakly (or not at all) with SM particles. The case of a massive U(1) gauge boson is colloquially referred to as a `dark photon.' 

The theory of a massive $U(1)$ gauge boson (`massive photon') dates back to 1936 work of Proca \cite{Proca:1936fbw}. In a modern context, gauge boson masses are understood to arise from either the Higgs mechanism or the Stuckleberg mechanism. In string theory, where $U(1)$ gauge bosons are ubiquitous, both the Higgs and Stuckleberg mechanisms are realized
\cite{Karozas:2020pun,Anchordoqui:2020tlp}.

The Proca action for a dark photon is necessarily but the first few terms in a low-energy effective field theory. One expects higher-derivative terms like $(F_{\mu \nu}F^{\mu \nu})^2$ and $(F_{\mu \nu}\tilde{F}^{\mu \nu})^2$ that appear in QED at energy scales below the electron mass (the famous Euler-Heisenberg Lagrangian \cite{Heisenberg:1936}), but also terms containing $A_\mu$, such as a quartic self-interaction $(A_\mu)^4$,  and derivative interactions such as $A_\mu \Box^2 A^\mu$. The Proca theory has a natural portal to the standard model, namely kinetic mixing with the Standard Model photon $F_{\mu \nu} ^{(\rm dark)} F^{\mu \nu {\rm (vis.)}}$ (see e.g., \cite{Holdom:1985ag,Abel:2008ai,Fabbrichesi:2020wbt}).

The dark photon effective field theory can also contain \textit{nonminimal} couplings to gravity, such as $R g^{\mu\nu}A_\mu A_\nu$ and $ R^{\mu\nu}A_\mu A_\nu$ (see e.g., \cite{Novello:1979ik,PhysRevD.31.1363,1972ApJ...177..757W,Toms:2015fja,Buchbinder:2017zaa,Ruf:2018vzq,Kolb:2020fwh, Ozsoy:2023gnl, Cembranos:2023qph,Capanelli2024}). These are dimension-4 operators consistent with the symmetries of the Proca theory, and thus should be included in the effective field theory. Moreover, even if neglected at tree-level, analogous to the $\xi \phi^2 R$ coupling of a scalar required for self-consistent quantization of a self-interacting field in curved space \cite{parker_toms_2009,birrell_davies_1982,TSBunch_1980}, the vector nonminimal couplings are expected as soon as self-interactions of $A_\mu$ are included, such as quartic terms $(A_{\mu})^4$, or from loop corrections to the the interaction vertex with gravity, namely the mass term $m_A^2 g^{\mu \nu }A_{\mu}A_{\nu}$ of Proca.

In this {\it Letter} we demonstrate that the nonminimal couplings, despite being perfectly consistent with symmetries of the Proca effective field theory, when considered in a cosmological context can induce a runaway production of arbitrarily high momentum particles, thereby causing a breakdown of the theory. We discuss the implication and interpretations of this in the context of different UV completions of the Proca theory.

{\it The Dark Photon in Curved Spacetime:}
On a generic spacetime background, $g_{\mu\nu}$, the Proca action can be written as \cite{Kolb:2020fwh}
\bea
S = & \int \dd^4x \sqrt{-g} \left(-\frac{1}{4}F^{\mu\nu}F_{\mu\nu} + \frac{1}{2}m^2 g^{\mu\nu}A_\mu A_\nu \right. \nn & - \left.\frac{1}{2}\xi_1 R g^{\mu\nu}A_\mu A_\nu - \frac{1}{2} \xi_2 R^{\mu\nu}A_\mu A_\nu\right), 
\label{eq:Sproca}
\eea 
where $A_\mu$ is the dark photon, $R$ the Ricci scalar, $R_{\mu\nu}$ the Ricci tensor and $\xi_1$ and $\xi_2$ dimensionless constants which couple the Proca field to gravity. The dark photon field strength, $F_{\mu\nu}$ is defined as $F_{\mu\nu} = \nabla_\mu A_\nu - \nabla_\nu A_\mu$. Notice that the interaction terms are the only dimension-four operators which can appear with the vector field coupled to curvature.
We would like to focus on cosmological production, so we now specialize to the spatially-flat Friedmann-Lemaitre-Robertson-Walker (FLRW) metric, $g_{\mu\nu} = a^2(\eta)\,\text{diag}(1, -1, -1, -1)$, where $a(\eta)$ is the scale factor as a function of conformal time.  In terms of the components $A_i$ and $A_0$, the action becomes 
\bea
S = & \int \dd^4 x\left[\frac{1}{2} a\left(\partial_0 A_i-\partial_i A_0\right)^2-\frac{1}{4} a^{-1}\left(\partial_i A_j-\partial_j A_i\right)^2\right. \nn
& \left.+\frac{1}{2} a^3 m_t^2 A_0^2-\frac{1}{2} a m_x^2 A_i^2\right], \label{eq:Scomponent}
\eea
with (time-dependent) effective masses, $m_t^2$ and $m_x^2$ as 
\begin{subequations} \label{eq:mtmx}
\begin{align}
    m^2_t = m^2 - \xi_1R - \tfrac{1}{2}\xi_2 R - 3 \xi_2 H^2 \label{eq:mt}\\
    m^2_x = m^2 - \xi_1R - \tfrac{1}{6}\xi_2R + \xi_2H^2, \label{eq:mx}
\end{align}
\end{subequations}
where $H$ is the Hubble parameter.  From \eref{eq:Scomponent}, we see that $A_0$ is not a dynamical variable and can thus be integrated out.  The next steps are standard \cite{Kolb:2020fwh}: Decompose the action in terms of mode functions, integrate out $A_0$, and introduce an orthonormal set of transverse and longitudinal mode functions.  The action then separates into two pieces.  Here we focus on the longitudinal component of the spin-1 field, with action \cite{Graham:2015rva,Ahmed:2020fhc,Kolb:2020fwh}
\bea
S^L & = \displaystyle\int \dd\eta \int \dfrac{\dd^3 \boldsymbol{k}}{(2 \pi)^3} \bigg[\tfrac{1}{2} \dfrac{a^2 m_t^2}{k^2+a^2 m_t^2} \left|\partial_0 A_{\boldsymbol{k}}^L\right|^2  \nn
&  -\frac{1}{2} a^2 m_x^2\left|A_{\boldsymbol{k}}^L\right|^2\bigg].
\eea
To ensure that the kinetic term is canonically normalized, we perform the field redefinition:
\be
A_{\boldsymbol{k}}^L(\eta)=\kappa_k(\eta) \chi_{\boldsymbol{k}}(\eta) \quad \text { with } \quad \kappa^2_k(\eta)=\frac{k^2+a^2 m_t^2}{a^2 m_t^2},
\label{eq:kappasq}
\ee
where we now suppress $L$ superscripts on $\chi$.
Notice that because $m_t^2$ is time dependent and not necessarily positive definite, $\kappa_k^2$ can potentially be negative, and a ghost can be propagated. If we demand a healthy theory that does not propagate ghosts, we must demand $\kappa^2(\eta)>0$. Requiring that this condition must be satisfied for arbitrarily large $k$ necessitates $m_t^2>0$.

The action for the longitudinal component is then 
\be \label{eq:Slong}
S^L  =\int\dd \eta \int \frac{\dd^3 \boldsymbol{k}}{(2 \pi)^3}\left(\tfrac{1}{2}\left|\partial_\eta \chi\right|^2-\tfrac{1}{2} \omega_k^2|\chi|^2\right) \, ,
\ee
where the longitudinal frequency is given by \cite{Capanelli2024}
\bea
\omega_k^2 & = k^2 \dfrac{m^2_x}{m^2_t} + a^2 m^2_x + \dfrac{3k^2a^4 m^2_t H^2}{(k^2 + a^2m_t^2)^2} + \dfrac{k^2a^2R}{6(k^2 + a^2m^2_t)} \nn
    & + \dfrac{H a k^2 m_t^{2'}}{m_t^2}\dfrac{(-k^2 + 2 a^2 m_t^2)}{(k^2 + a^2m_t^2)^2}  - \dfrac{k^2 m_t^{2''}}{2m_t^2(k^2 + a^2 m_t^2)} \nn
    &  + \dfrac{k^2(m_t^{2'})^2}{4 (m_t^2)^2}\dfrac{(k^2 + 4 a^2 m_t^2)}{(k^2 + a^2m_t^2)^2} \ ,
    \label{eq:omegaL}
\eea
where prime denotes $\partial_\eta$ and the wavenumber $k=|\boldsymbol{k}|$. The action for the transverse component is similar to Eq.\ \eqref{eq:Slong} but with $\omega_k^2=k^2+a^2m_x^2$. In each case the equation of motion of the mode function $\chi$ is given by $\chi_k ''+ \omega_k ^2 \chi_k=0$.

In the high momentum (large-$k$) limit, the longitudinal frequency, Eq.~\eqref{eq:omegaL}, is dominated by the first term, $\omega_k^2 \to k^2 m^2_x/m^2_t$.  The evolution of high-momentum modes is therefore dictated by the evolution of the effective sound speed $m^2_x/m^2_t$. 
Since we have established that $m^2_t>0$ for a ghostless theory, it follows that if $m_x^2<0$, then $\omega_k^2$ will be \textit{negative}, leading to an instability to particle production of arbitrarily large $k$ modes. This is similar but distinct from the instability of spin-3/2 particles observed in \cite{Kolb:2021nob,Kolb:2021xfn}. In what follows, we refer to this phenomenon as {\it runaway production}.

{\it Representative cosmological model:}
For numerical examples of runaway production we assume an inflationary model with a quadratic potential.  The salient features of this model are common to a wide range of inflationary models, namely, a quasi-de Sitter (qdS) phase followed by a matter-dominated (MD) phase driven by the coherent oscillations of the inflaton field. For an FLRW cosmology with fixed equation of state $w$ [i.e., ignoring any oscillations], we have $r\equiv R/H^2=-3 + 9w$.  In the qdS phase $r\simeq -12$ and in the MD phase $r$ oscillates between $r=-12$ (when the inflaton field is at the maximum of the potential and momentarily in a de Sitter phase with $w=-1$) and $r=6$ (when the inflaton is at the minimum of the potential and momentarily in a kination phase with $w=+1$).  The average value of $w$ in the MD phase is zero.

Let us first examine the conditions for a ghostless theory, $m_t^2\geq 0$, in the limit $m/H \rightarrow 0$. (Dark photons of recent phenomenological interest \cite{Caputo_2021,Heeba:2023bik,Dolan:2023cjs,Brahma:2023psr} are  $\mathcal{O}(10)\,{\rm MeV}$, while $H$ is as large as $10^{13}$\, GeV, making $m\ll H$ a well motivated assumption).  In this case the ghostless requirement is $-r(\xi_1+\tfrac{1}{2}\xi_2)-3\xi_2\geq 0$.   Using $-12<r<6$, this leads to $-\xi_2 > \xi_1 > -\xi_2/4$.  This can only be satisfied if  $\xi_2<0$.  Now, consider the requirement $m_x^2>0$ to avoid runaway particle production: $-r(\xi_1+\tfrac{1}{6}\xi_2)-\xi_2>0$.   Runaway will be avoided if $0>\xi_1>-\tfrac{1}{4}\xi_2$.  This is impossible to satisfy for $\xi_2<0$.  Thus, in the limit $m\ll H$, the requirements of ghostless and no runaway are incommensurate unless $\xi_1=\xi_2=0$. 

\begin{figure}
    \centering    
        \includegraphics[width=0.48\textwidth]{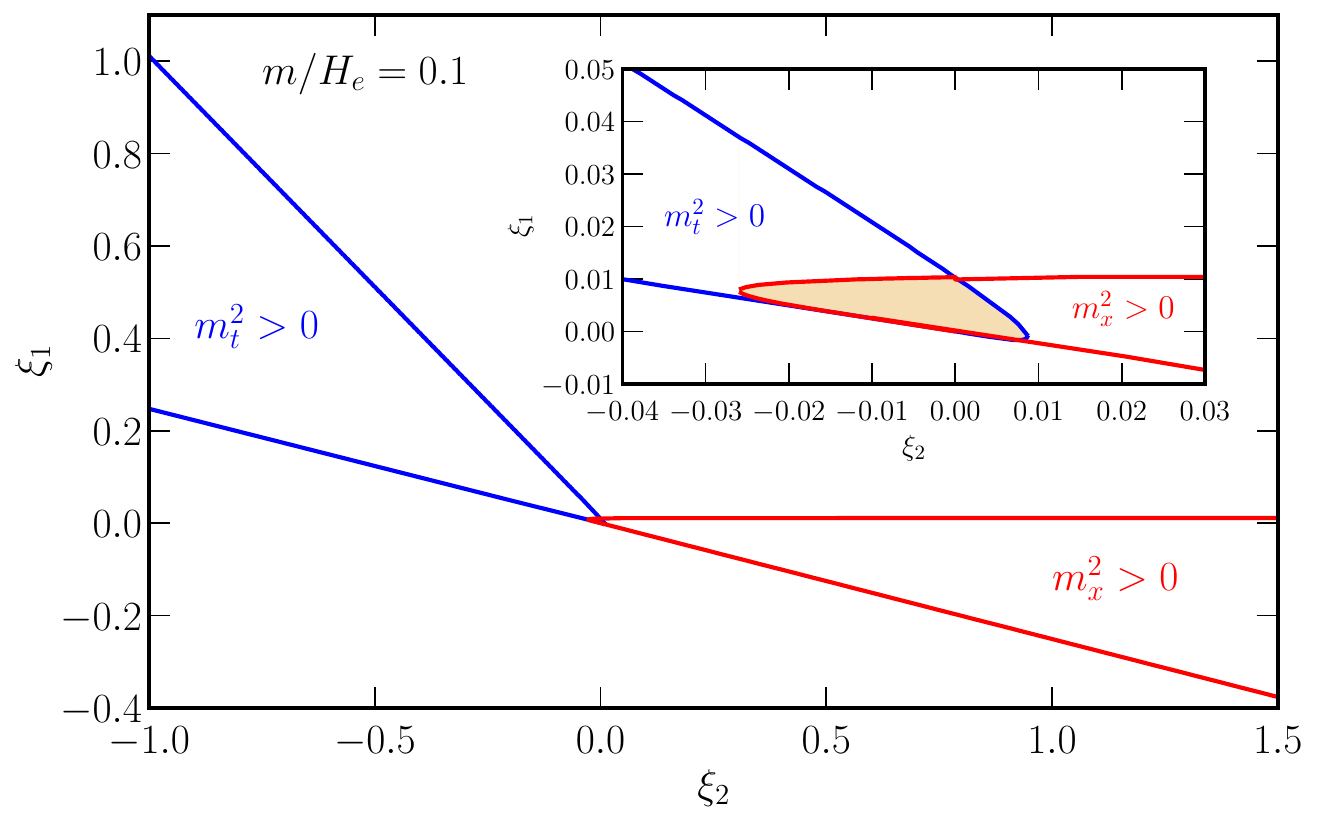} \\
        \includegraphics[width=0.48\textwidth]{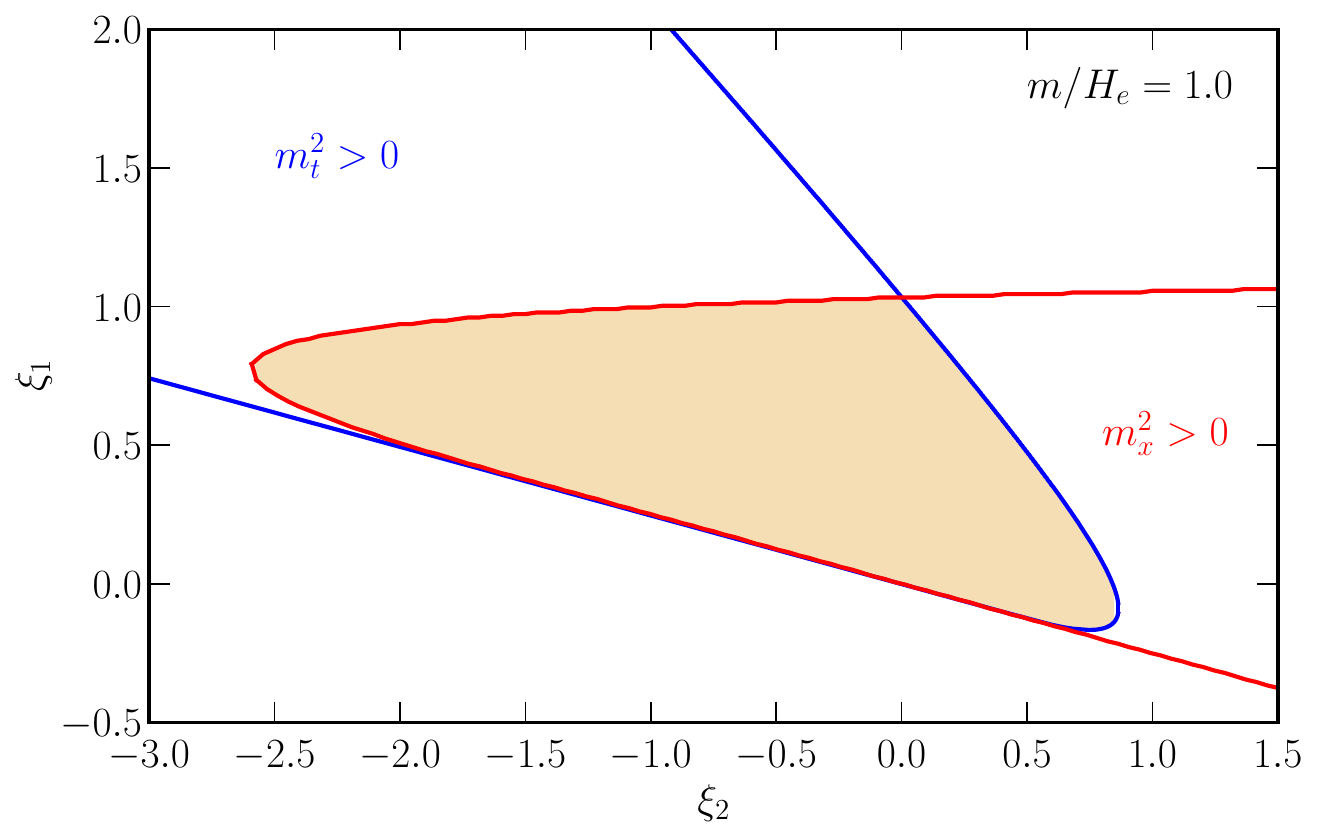}
   \caption{
    In the $\xi_2-\xi_1$ plane, the interior of the blue contour has $m_t^2>0$, hence ghostless, and the interior of the red contour has $m_x^2>0$, hence the intersection of the two regions has no runaway GPP.  The shaded region is ghostless and GPP is well-behaved for large $k$. The upper panel is for $m/H_e=0.1$, while the lower figure is for $m/H_e=1$.  The insert in the top panel is a blow-up of the region surrounding $(\xi_2,\xi_1)=(0,0)$ showing the small region without ghost/runaway.}
    \label{fig:xi1xi2}
\end{figure}

For finite $m/H$ it is possible to find values of $(\xi_2,\xi_1)$ that are both ghostless and runaway safe.  Clearly from \eref{eq:mtmx} for sufficiently large $m$ the right hand side of the equations will be positive.  This is illustrated in Fig.\ \ref{fig:xi1xi2} for two choices of $m/H_e$, where $H_e$ is the Hubble parameter at the end of inflation.  The choice $m/H_e=0.1$ approximates the $m/H\to0$ case and there is just a small shaded region where the theory is ghostless and UV-safe.  As $m$ is increased, the safe region grows.

\begin{figure}
    \centering
    \includegraphics[width=0.48\textwidth]{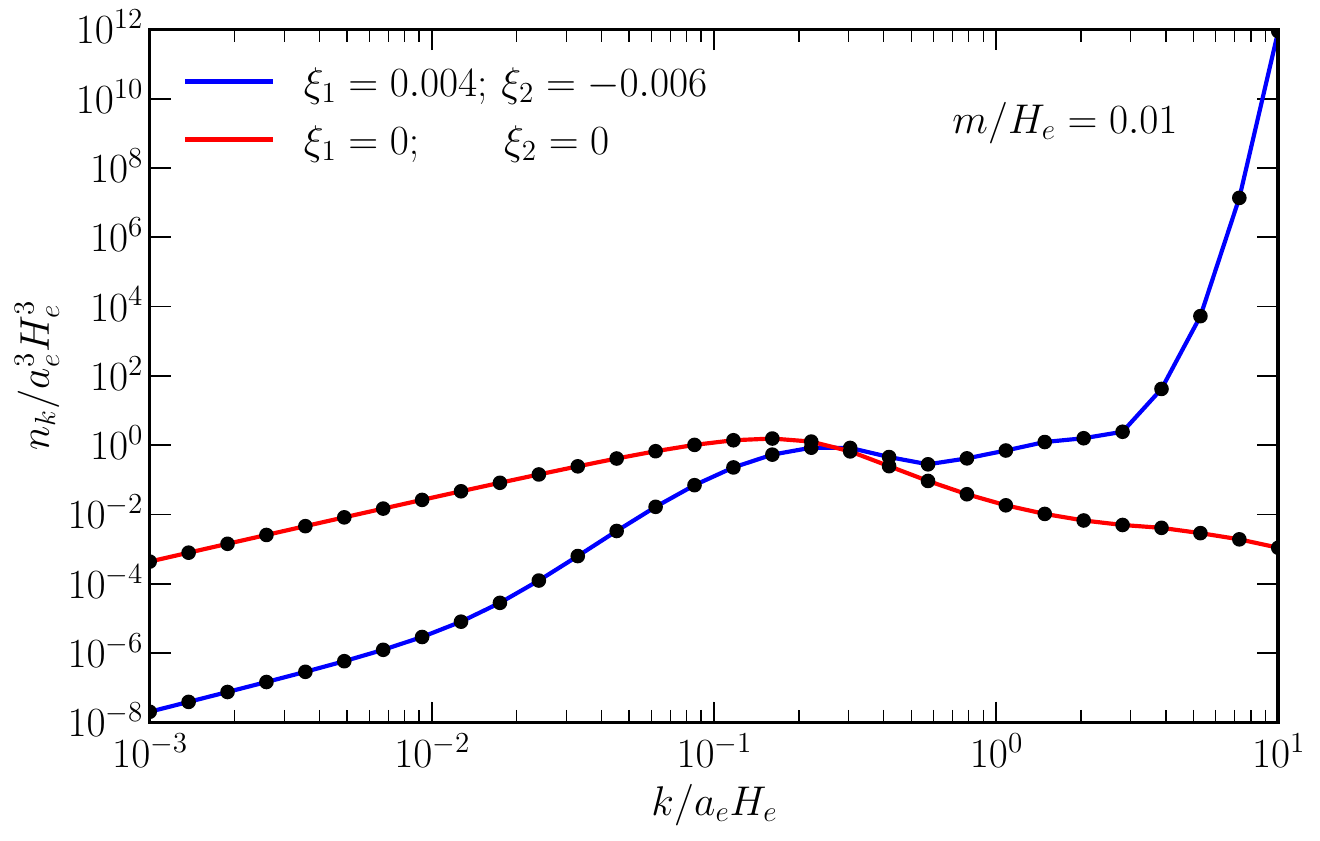} 
    \caption{The spectral density $n_k/a_e^3H_e^3$ as a function of wavenumber $k/a_eH_e$ for the longitudinal mode for two choices of $(\xi_1,\xi_2)$, with $m/H_e=0.01$ in both cases.  The choice $(0,0)$ is minimal coupling; the nonminimal choice $(0.004,-0.006)$, while ghostless, has $m_x^2<0$.  The minimal choice results in $n_k$ well behaved at high $k$, while the nonminimal choice leads to runaway production at large $k$.}
    \label{fig:runaway}
\end{figure}

{\it Runaway Particle Production.} 
We numerically solve for the evolution of the mode functions $\chi_k$ and from this compute the comoving number density of particles $n_k=k^3|\beta_k|^2/2\pi^2 $ with $  |\beta_k|^2 = \omega_k|\chi_k|^2/2 + |\partial_\eta \chi_k|^2/2\omega_k-1/2$ (see \cite{Kolb:2023ydq}).
In Fig.~\ref{fig:runaway} we show the result for the longitudinal components of a vector field of mass $m=0.01\ H_e$, where $H_e$ is the expansion rate at the end of inflation, for two values of $(\xi_1,\xi_2)$: $(0,0)$-minimal gravitational coupling, and a model with nonminimal couplings, $(0.004,-0.006)$. The nonminimal parameters were chosen such that the model is ghostless ($m_t^2>0$) but has a high-$k$ runaway ($m_x^2<0$).  Here, the physical momentum of a mode is $k/a$, so $k/a_e$ is the physical momentum of the mode at the end of inflation.  Modes with $k/a_eH_e>1$ were inside the Hubble radius at the end of inflation. The total number density of particles due to GPP is $na^3=\int n_k\,{\rm dlog}k$ (see \cite{Kolb:2023ydq} for details).  

From Fig.~\ref{fig:runaway} one may appreciate a dramatic amplification of the nonminimally coupled model (blue) at high-$k$. Compared to the minimal coupling model (red), the nonminimal model has $n_k/a_e^3H_e^3$ $40$ times larger at $k/a_eH_e =1$ and $10^{15}$ times larger at $k/a_eH_e=10$. This exemplifies runaway production.

Clearly there is an issue if $n_k$ does not turn over for large-$k$: the Proca theory is itself an EFT, and production of modes at or above the cutoff would necessarily cause a breakdown of the EFT description.  The simplest way out is therefore posit a UV cutoff $\Lambda$ above which production is tamed. To this end, it is useful to relate a physical momentum $\Lambda$ to a comoving wavenumber $k$: $\Lambda = k/a$.  For the example illustrated for $m/H_e=0.1$ in Fig.\ \ref{fig:runaway},  the angular frequency $\omega_k^2$ first becomes negative for high-k modes around $ a/a_e\sim 1$.  Simply taking $a/a_e=\mathcal{O}(1)$, strictly positive $\omega_k^2(k)$  would require a physical cutoff $\Lambda\lesssim H_e$. This is clearly untenable; it would invalidate the whole analysis of quantum fluctuations during inflation.

Either something within the Proca EFT must regularize the high-$k$ behavior, or else there must be a value of $k$ beyond which our calculation is invalid.

We will now discuss three possibilities:  1) the UV completion of the Proca theory could cure the high-$k$ runaway; 2)  kinetic mixing with the standard model photon may resolve the instability; 3) the Proca theory becomes strongly coupled at some momentum scale, and beyond this scale our results can not be trusted.

{\it Dark photon effective field theory:} 
The UV physics is encoded into the EFT by nonrenormalizable operators including derivative terms like $A_\mu \Box^2 A^\mu /\Lambda^2$ which would contribute to $\omega_k^2$ a term like $k^4/\Lambda^2$. One might hope this might shutoff the instability while maintaining validity of the EFT, namely that the conditions (1) that $k/\Lambda > |m_{x}/m_t|$ (to make $\omega_k^2$ positive at high $k$) and (2) $k/\Lambda \ll 1$ to maintain perturbative control, can be satisfied simultaneously. However, there is no hierarchy between $m_{x}$ and $m_{t}$: $m_{x}/m_t$ oscillates to ${\cal O}(1)$ negative values in the instability region. Concretely, for $m\ll H$, one can consider three regimes: $\xi_1 \gg |\xi_2|$, which gives $m_x/m_t \rightarrow 1$,  $\xi_1 \ll |\xi_2|$ which gives $|m_x/m_t|\rightarrow 1/\sqrt{3}$, or $\xi_1 \sim |\xi_2|$ which gives $|m_x/m_t|=\sqrt{|5/3 + 8/(6-r)|} = {\cal O}(1)$. Thus, it is not possible for higher-derivative terms in the EFT to shutoff the instability at high-$k$ while maintaining perturbative control in the EFT.  It follows that any hope of resolving the runaway lies in abandoning the EFT in favor of an explicit UV completion.

{\it UV Completion:} 
To make more progress we can turn to an explicit UV completion. As a simple UV example we consider the Abelian Higgs model, with Lagrangian 
\begin{equation}
    {\cal L}=-\frac{1}{4} F_{\mu \nu} F^{\mu \nu} + |D_\mu \phi|^2 + \mu^2 |\phi|^2 - \lambda |\phi|^4
\end{equation}
where $\phi$ is a complex scalar, and $D_{\mu}$ is the gauge covariant derivative $D_\mu=(\partial_\mu - i g A_\mu)$ where $g$ is the gauge coupling. We expand $\phi$ about the minimum at $\phi= \mu/\sqrt{\lambda}\equiv \sqrt{2} v$ as  $\phi = (v + h)e^{i\theta}/\sqrt{2}$, where $h$ is a real scalar and fix the $U(1)$ gauge symmetry to unitary gauge where $\theta=0$ and the photon is massive. The Lagrangian is then simplified to ${\cal L}=\frac{1}{4} F_{\mu \nu} F^{\mu \nu} + \frac{1}{2}(\partial_\mu h)^2 + \frac{1}{2}m_{A}^2 A_{\mu}A^{\mu}\left( 1 + \frac{h}{v}\right)^2 + \frac{1}{2} m_h^2 h^2 + 4 v \lambda h^3 + \lambda h^4$, corresponding to a Higgs mass $m_h=\sqrt{\lambda/2}\,v$, a photon mass of $m_A = g v/2$, and cubic and quartic interactions between $h$ and $A_\mu$ given by \cite{Peskin:1995ev}
\begin{equation}
    {\cal L}_{AAh}=  \frac{m_{A}^2}{v} A_{\mu}A^{\mu} h \,\, , \,\,   {\cal L}_{AAhh} = \frac{m_A^2}{v^2}A_{\mu}A^\mu h^2.
    \label{eq:ints}
\end{equation}
Meanwhile, the gravitational couplings of the gauge field generate interactions with gravitons
which at cubic order in  the fields (after performing a series expansion of $\sqrt{-g}Rg^{\mu\nu}$ and $\sqrt{-g}R_{\mu \nu}$) are given by
\begin{eqnarray}
    {\cal L}_{AAg} = \frac{\sqrt{-g}}{M_{pl}}\left( m_A^2 + \xi_1 R + \xi_2 D_{\alpha}D^\alpha \right) \delta g^{\mu \nu}A_{\mu} A_{\nu} ,\nonumber
\end{eqnarray}
where $\sqrt{-g}$, $R$, and $D$ are defined with respect to the background geometry, $\delta g_{\mu \nu}$ denotes a canonically normalized transverse traceless fluctuation to the metric of mass dimension 1 ($g_{\mu \nu} = g_{\mu \nu} ^{(0)} + \delta g_{\mu \nu}/M_{pl}$).  There are additional interactions at quartic order, e.g., from the expansion of $\sqrt{-g} R$, of the form ${\cal L}_{AA gg}\sim (m_A^2 + \xi_1 D^2)  \delta g_{\mu \nu} \delta g^{\mu \nu} A_\sigma A^\sigma /M_{pl}^2+...$\,.

Let us consider the possibility that the nonminimal gravitational coupling constants $\xi_1$ and $\xi_2$ vanish at tree-level, and are generated as effective interactions via loops in the Abelian Higgs model. With the interactions between $h$ and $A$ given by Eq.~\eqref{eq:ints}, loops of $\phi$ particles renormalize the minimal cubic interaction of $A_{\mu}$ with gravity, which by dimensional analysis generates a nonminimal coupling with $\xi_{2}\propto (m_{A}/m_h) ^6$. Similarly, the 1-loop correction to the minimal quartic interaction generates $\xi_1 \sim (m_A/m_h)^6$. Depending on the relative size of the Higgs and gauge couplings, $\lambda$ and $g$ respectively, the effective couplings $\xi_{1,2}$ can be made small, but are not necessarily so. Meanwhile, the instability depends on the ratio $m_A/H$, with $H$ the Hubble parameter, whereas the expected size of $\xi_{1,2}$ is independent of $H$. Thus even the ostensibly small loop-generated couplings can lead to a runaway production: UV completion into the Abelian-Higgs model does not in itself regulate the instability. 

The loop-induced couplings present a similar situation to that with nonminimally coupled scalars: while the coupling may be set to zero at one energy scale, renormalization group (RG) flow will generate non-zero value of the coupling at all other energy scales. This justifies the Wilsonian intuition that all terms allowed by symmetries should be allowed in the low-energy effective field theory, and absent any measurement to anchor the RG flow, the coupling constants should be taken to be free parameters. 

With this in mind, we can include the vector nonminimal couplings directly in the Abelian-Higgs model, where they manifest as nonminimal derivative couplings,  ${\cal L}_{1} =  \frac{\xi_1}{(gv)^2} R D_{\mu}\phi D^\mu \phi^*$ and  ${\cal L}_2=  \frac{\xi_2}{(gv)^2}D_{\mu}\phi D_{\nu }\phi^* R^{\mu \nu} $.
This provides a gauge-invariant UV completion of Eq.~\eqref{eq:Sproca}.  Couplings of this form have been extensively studied for a neutral scalar field such as \cite{Amendola:1993uh}. The vector nonminimal couplings now modify the kinetic action of the Higgs scalar field. One may easily appreciate that $\xi_{1,2} \neq 0$ can lead to ghost and/or tachyonic instabilities, just as in the dark photon model that emerges at low energies. Thus again one sees that the UV completion is not the cure.

{\it Kinetic Mixing with the Standard Model:} 
Another possibility for curing the runway, within the dark photon EFT, is dark-photon interaction with the Standard Model (SM), namely the kinetic mixing ${\cal L}_{int}=\epsilon F_{\mu \nu} ^{(\rm dark)} F^{\mu \nu {\rm (SM)}}$. In the case that the dark photon is massive (as we study here), the kinetic terms can be diagonalized via the field redefinitions \cite{Fabbrichesi:2020wbt}    $A_{\mu} ^{\rm SM}\rightarrow  A_{\mu} ^{\rm SM'}\equiv A_{\mu} ^{\rm SM} - \epsilon A_{\mu} ^{\rm dark}$ and $ A_{\mu} ^{\rm dark}\rightarrow A_{\mu} ^{\rm dark} $. The resulting action has canonical kinetic terms. This endows SM fields with dark-photon charge, i.e., interactions of the form $A_{\mu}^{\rm dark}\bar{f}\gamma^\mu f$. The nonminimal gravitational terms are unaffected. Thus, in the diagonal basis, the only change generated by the kinetic mixing is the addition of charged fields in the Proca theory. At tree-level, the charged fermions do not alter the kinetic sector of the dark photon, and thus have no impact on the $k^2$ culprit of the runaway production. On the other hand, loops of charged fields can generate higher-derivative terms in the Proca EFT, such as $k^4$ terms, but these fail to cure the runaway, as described above. Thus kinetic mixing does not prevent or preclude the runaway production.

{\it Strong coupling:}
The Proca theory nonminally coupled to gravity should become {\it strongly coupled} at some scale. Lacking a first principles derivation of the strong-coupling scale we can still consider the implications of the existence of such a scale. Similar but distinct from a breakdown of the EFT, strong coupling would render the evolution of high-$k$ modes impervious to calculation. In practice, the consequences for dark photon phenomenology remain unchanged from the weakly coupled runaway, namely conventional tools of dark matter phenomenology (such as perturbative QFT) no longer apply.


{\it Conclusions:} 
In this work we have pointed out the possibility of runaway cosmological production of dark photons with nonminimal couplings to gravity. We studied this phenomenon in the context of inflation with a quadratic potential and late reheating, however we emphasize that the discussion above is largely generic and independent of both the inflation model and reheating. This is discussed in further detail in \cite{Capanelli2024}, in which we show that the particle production remains qualitatively similar in scenarios with rapid-turn multifield inflation (see also Ref.~\cite{Kolb:2022eyn}) and/or early reheating. Also, in \cite{Capanelli2024} we point out that even if high-$k$ runaway is stopped at some moderate value of $k/a_eH_e$, values of $(\xi_1,\xi_2)$ where $m_x^2<0$ have very much enhanced GPP compared to the minimal model  For instance, from \fref{fig:runaway} we see that for $k/a_eH_e=10$ GPP in the nonminimal model is enhanced by about a factor of $10^{15}$ compared to the minimal model, even for small values of $(\xi_1,\xi_2)$. If the runaway production can be tamed, while leaving an overall enhanced production, this effect could widen the range of parameters to result in dark photons as dark matter. Finally, it will also be interesting to examine whether the runaway production also exists for nonminimally coupled pseudovector fields, as realized by the Kalb-Ramond theory of an antisymmetric tensor field, see Ref.~~\cite{Capanelli:2023uwv} for Kalb-Ramond GPP.

{\it Acknowledgements:} 
The authors thank B.\ Grzadkowski and A.\ Socha for communication of unpublished work and useful comments on a preliminary version of this paper, and A.\ Hell for information about strong coupling in Proca theories.  C.C.\ acknowledges support from the Canadian Institute for Particle Physics (IPP) via an Early Career Theory Fellowship.  C.C.\ is supported by a fellowship from the Trottier Space Institute at McGill via an endowment from the Trottier Family Foundation, and by the Arthur B. McDonald Institute via the Canada First Research Excellence Fund (CFREF). The work of L.J.\ is supported by the Kavli Institute for Cosmological Physics at the University of Chicago. The work of E.W.K.\ was supported in part by the US Department of Energy contract DE-FG02-13ER41958 and the Kavli Institute for Cosmological Physics at the University of Chicago.  E.M. is supported in part by a Discovery Grant from the Natural Sciences and Engineering Research Council of Canada, and by a New Investigator Operating Grant from Research Manitoba.

\bibliography{letter}
\bibliographystyle{JHEP}

\end{document}